**Network Dynamics Governed by Lyapunov Functions: From Memory to Classification**

Merav Stern[1], Eric Shea-Brown[1,2]

[1] Department of Applied Mathematics, University of Washington, Seattle, WA, 98195-3925, USA

[2] Allen Institute for Brain Science, Seattle, WA, 98109-4307, USA

Abstract:

In 1982 John Hopfield published a neural network model for memory retrieval, a model that became a cornerstone in theoretical neuroscience. A key ingredient of the Hopfield model was the use of a network dynamics that is governed by a Lyapunov function. In a recent paper, Krotov and Hopfield showed how a Lyapunov function governs a biological plausible learning rule for the neural networks' connectivity. By doing so, they bring an intriguing approach to classification tasks, and show the relevance of the broader framework across decades in the field.

Text:

The Hopfield model [1] has proven one of the most influential theoretical models in neuroscience; for many researchers, it also served as a cornerstone in their introduction to the computational neuroscience world. At its core lies the idea that a pattern of sustained neural activity can represent a memory, which is retrievable via a network dynamics that converges into that memory. The model is defined by a connected network of neurons whose activity follows a dynamical update equation [2, 3]. To allow convergence of the network into stable memories, Hopfield used symmetric connectivity. This property facilitated analytical calculation of the memory capacity of the network, an approach which paved the way for many other results (see for example [4]).

In the model, a memory is represented as a pattern of sustained activity in the network. The activity of each element in the network, a neuron, can have two possible values representing it being either active (spiking) or silent (not spiking). The ongoing activity of each neuron depends on its input, which results from the network activity at the previous time step and the weights of connections from other neurons. The memories are imprinted into the network via its symmetric connectivity structure. As a result, each memory in this model is a stable state of the network, and by following its dynamics the network has a better chance to converge to a certain memory as the correlation between the network state and the memory increases.

A crucial insight made by Hopfield is that the resulting model dynamics are governed by a Lyapunov function (also refers to as the energy function) – a mathematical function of the network activity that monotonically decreases whenever the state of the network changes and yet is bounded from below. Hence, a network with a Lyapunov function descends along the function's value to a local minimum. In the Hopfield model, these minima are governed by the memories imprinted into the network, as illustrated in Figure 1A. Importantly, this property is shared by more general continuous networks as well [5, 6].

A recent paper by Krotov and Hopfield [7] presents another possible advantage for neural network dynamics that follows a Lyapunov function, this time to solve a classification task. In their new study, Krotov and Hopfield built a three layer feedforward neural network, as illustrated in Figure 1B. The network connectivity between the input layer and the hidden (middle) layer dynamically changes in response to the activity driven by the inputs (unsupervised learning), following dynamics governed by a Lyapunov function. Later, the network connectivity between the hidden layer and the output layer is trained to classify the input patterns according to their labels (supervised learning). The network incorporates biologically plausible learning (as in, for example, [8-12]). Specifically, each of its connection weights between the input layer and the hidden layer follows a Hebbian-like plasticity rule [13, 14] while being bounded by constraining the vector length of all weights.

Hopfield and Krotov's choice for the boundary between activity strength that drives potentiation and depression in their implementation of Hebbian plasticity follows an intriguing rule. This boundary is chosen by examining the activity of the hidden layer, as driven by all the input examples, leaving potentiation to drive only a few hidden units at a time while depressing others (and not updating connections where signs of inputs and weights are opposite).

In the original Hopfield model discussed earlier, a large correlation value between the network and a memory state pushes the network farther towards that memory while suppressing other, uncorrelated memories. The more recent model by Krotov and Hopfield uses a similar mechanism. Their choice to decrease weights (synapses) that are connected to a weakly activated neuron, for the majority of neurons in the hidden layer that don't cross their chosen boundary, causes patterns that strongly activated neurons in the hidden layer to push these neuron's synapses even farther towards these patterns, while depressing synapses of neurons that were only weakly activated by these patterns.

The network follows a Lyapunov function defined by the integral of the function that sets the boundary between potentiation and depression. In the context of the Krotov and Hopfield network, this means that the dynamics lead this function to monotonically change until a local maximum has been reached, as illustrated in Figure 1B. Hence the network maximizes the responses of the hidden layers to inputs within the normalization constraints, a productive distribution of its weights.

When the training described above is complete, the activity of the hidden units is fed into an output layer of a perceptron, which is trained using supervised learning (via stochastic gradient decent). Krotov and Hopfield show that this scheme can perform classification, of handwritten digits (MNIST) for example, with high accuracy.

The success of Krotov and Hopfield in showing how a biologically plausible learning rule can be governed by a Lyapunov function opens new doors to understanding how networks learn to perform classification, among other fundamental tasks. This is important, as efforts to analyze the underlying mechanisms are central to both neuroscience and artificial intelligence.  It is also inspiring, as it comes almost forty years after Hopfield's original discovery of how Lyapunov functions can govern the dynamics of memory.  Noting how this laid groundwork for so many of the studies that were to follow, we are keenly looking forward to a deepened understanding of classification and other computations in biological plausible networks in the years ahead.

Acknowledgements:

We acknowledge support from the Sackler Foundation, the Swartz Foundation via the Swartz Center for Theoretical Neuroscience at the University of Washington, the NSF ERC Center for Sensorimotor Neural Engineering, and NSF DMS Grant #1514743. We thank the Allen Institute founders, Paul G. Allen and Jody Allen, for their vision, encouragement and support.

Network Dynamics Governed by Lyapunov Functions: From Memory to Classification    M. Stern and E. Shea-Brown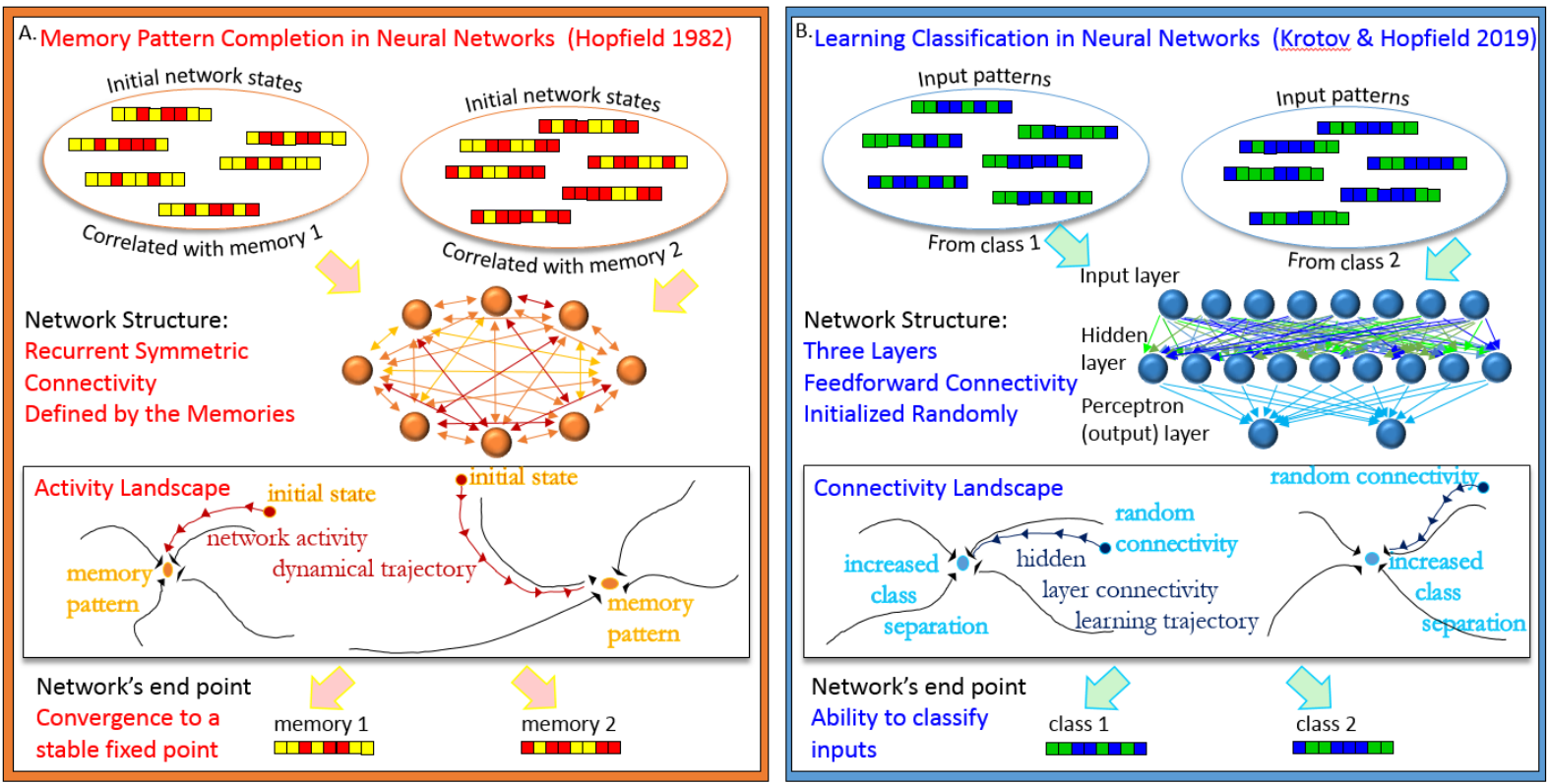

Figure legend

Figure 1: Structure and activity of neural networks with Lyapunov functions.
A (left orange box). Hopfield neural network model [1] for memory retrieval.
B (right blue box). Krotov and Hopfield neural network model [4] for classification.
Top ellipsoids: Network initializations. The schematic illustrates two sets of six possible initializations for each of the networks. Each initialization is given by a vector with binary values, indicated by two possible colors, and representing each neuron being in an active (spiking) or in a silent (not spiking) state.
Left, Hopfield network: Each element in the vector defines one neuron's initial state for each of the eight neurons in the network. All initial states in the right and left ellipsoids are highly correlated with memory state 1 and memory state 2, respectively (drawn at the bottom).
Right, Krotov and Hopfield network: Each element in the vector defines an input pattern to one of the eight neurons in the input layer of the network. Here input values are continuous, rather than discrete, and can be any real number. All input patterns within a set are highly correlated with an example that represents their class (these examples are drawn at the bottom).
Middle: Network structures. Each node represents a neuron in the network. Each arrow represents an input from one neuron to another, forming together the network connectivity structure.
Left, Hopfield network: The network connectivity is symmetric (evident by the dual direction arrows). The connectivity structure is fully defined by the memories.
Right, Krotov and Hopfield network: The network is composed of three layers, with feedforward connectivity between the layers, as the single-directional arrows indicate.
Middle box: Network dynamics. The Lyaponuv function landscape is illustrated with two attractors, each a local minimum of the function. The dynamics are attracted towards these, which serve as local stable states of the network.

Left, Hopfield network: Each local minimum corresponds to a memory state. The network starts in an initial state and follows the dynamics. Here, each neuron is updated according to the input it receives from other neurons, defined by the network activity at the previous time step and the weights of connections. This dynamic descends along the Lyaponuv function's value, as indicated by the red examples of possible trajectories, to a local minima, indicated by orange circles, representing memories or stable states in the network.

Right, Krotov and Hopfield network: Each local minimum corresponds to a set of connection weights between the input and the hidden layer. The network starts in a given connectivity structure and the connectivity weights between the input layer and the hidden layer follow the dynamics at hand. Here, weights change their values in response to the activity driven by the inputs, according to Hebbian like plasticity rule that is constrained by the vector length of all weights. This dynamic ascends along the Lyaponuv function's value, as indicated by the blue examples of possible trajectories, to a local maximum, indicated by light blue circles, representing attracting connectivity structures. These are stable states of the network connectivity.

Bottom: End result of the network.

Left, Hopfield network: The network final steady state can be memory 1 or memory 2, according to the initial state with which it had higher correlation.

Right, Krotov and Hopfield network: The network final structure allows to classify inputs according to the class with which the input had higher correlation.